\documentstyle[multicol,aps,epsfig]{revtex}

\begin{document}

\newcommand{\wt} {\widetilde}

\title{Andreev reflections in the
pseudogap state of cuprate superconductors
}

\author{Han-Yong Choi,$^{1,2}$ Yunkyu Bang,$^{3}$
and David K. Campbell$^{2,4}$  }
\address{$^{1}$Department of Physics and
Institute for Basic Science Research,\\
Sung Kyun Kwan University, Suwon 440-746, Korea.}
\address{$^{2}$Loomis Laboratory of Physics, University of Illinois
at Urbana-Champaign, \\
1110 West Green Street, Urbana, IL 61801.}
\address{$^{3}$Department of Physics, Chonnam National University,
Kwangju 500-757, Korea.}
\address{$^{4}$CNLS, Los Alamos National Laboratory, Los Alamos,
NM 87545.}

\date{\today}

\maketitle

\begin{abstract}
We propose that, if the pseudogap state in the cuprate superconductors can
be described in terms of the phase-incoherent preformed pairs,
there should
exist Andreev reflection from these pairs even above the
superconducting transition temperature, $T_c$.
After giving qualitative
arguments for this effect, we present more quantitative calculations based on
the Bogoliubov--de Gennes equation. Experimental observations of the
effects of Andreev reflections above
$T_c$---such as an enhanced tunneling conductance below the gap along
the copper oxide plane---could provide unambiguous evidence for the preformed
pairs in the pseudogap state.
\end{abstract}


\begin{multicols}{2}

Underdoped to slightly overdoped cuprate superconductors exhibit
a progressive transfer, as the temperature is lowered,
of the low-energy spectral weight to the higher energy region
in both the spin and charge channels \cite{timusk,randeria}.
The resulting ``pseudogap'' begins to appear at a cross-over temperature
$T^*$, well above the superconducting transition temperature, $T_c$,
and the nature of this pseudogap state has been the focus of
considerable recent interest.
Many proposals have been presented to explain this
phenomenon, including
various forms of electron pairing \cite{emery,levin,randeria2},
antiferromagnetic fluctuations \cite{pines,chubukov},
and spin-charge separation scenarios
\cite{lee,anderson}, but no consensus on the origin of the pseudogap
has emerged. Among the
electron pairing scenarios, the phase-incoherent preformed pair (PIPP), or,
normal state pair, scenario suggested by Emery and Kivelson \cite{emery} seems
particularly intriguing, since the amplitude of the pairing order parameter is
non-vanishing even above $T_c$. In this scenario, the PIPP begin to
appear at $T^*$, but do not produce superconductivity until $T_c$ because the
thermally excited phase fluctuations of the pairing order parameter destroy
phase coherence and hence superconductivity for $T>T_c$.
%Recent experiments provide seemingly conflicting data
%regarding this scenario. 
Several earlier experiments,
as well as recent teraherz spectroscopy results \cite{orenstein},
seem to support the idea of the PIPP,
but their existence has not been established \cite{timusk,pennington}.
%However, recent NMR measurements \cite{pennington},
%which find that $T_c$ is decreased by as much as 8 K by a
%magnetic field of 14.8 T while $T^*$ is hardly changed at all,
%appear to cast doubt on the PIPP scenario.
In the present paper, we show that a
crucial test of the PIPP scenario will be the possible
Andreev reflection (AR) \cite{andreev,bruder} from
the pseudogap state: We propose that if the pseudogap state contains
PIPP, AR will occur and will produce,
among other effects, an enhanced conductance even {\it above
$T_c$} that should be readily measurable in tunneling conductance
experiments.

Two qualitative intuitive arguments suggest that AR should
occur from a state containing PIPP.
First, AR (or ``retroreflection of a hole'') in a normal
metal--superconductor (NS) junction occurs because an electron, which
is a well-defined elementary excitation (quasiparticle) in the normal
metal, is $not$ an elementary excitation of the superconductor.
Therefore, an electron from the normal metal, upon entering the
superconductor, has to ``reconstruct'' itself as a linear combination of the
quasiparticles of the superconductor, namely, the Bogoliubovons.
This requires an electron of momentum ${\bf k}$ and spin $\sigma$ to
find a time-reversed mate electron of momentum $- {\bf k}$
and spin $- \sigma$ to form a pair. The hole left behind by the mate
electron then retraces the path of the incoming electron, which is the
AR. Naively, the same underlying physics
should apply to the (assumed) preformed pairs in the pseudogap state,
because an electron is not an elementary excitation of that
state either, once the pairs are formed, {\it independent} of their phase
coherence. The second heuristic argument involves the phase stiffness.
The superfluid density in the static, long
wavelength limit, $\rho_s (T)$, which also determines the
superconductor's phase stiffness, $\Theta(T) =\hbar^2 \rho_s(T)/k_B$,
vanishes for $T>T_c$.
In the dynamic  Kosterlitz-Thouless-Berezinsky (KTB) theory,
$\rho_s (\omega,T) \sim \omega \tau_c/(1+\omega\tau_c) >0$ as
$\omega$ is increased from 0, where $\tau_c$ is the coherence time
\cite{minnhagen}.
Precisely this finite frequency
phase stiffness, measured by a teraherz spectroscopy,
was recently reported to scale as $\Theta_{KTB} = (8/\pi) T_{KTB}$,
in accord with the KTB theory
of the classical phase fluctuations \cite{orenstein}.
$T_{KTB} = T_c$ for two-dimensional (2D) superfluids.
Importantly, the superfluid density {\it at a finite
wavevector} (and $\omega =0$), $\rho_s (q,T)$ does
not vanish identically above $T_c$ but is given
by $\rho_s (q,T) \sim (q \xi_+)^2$, where $\xi_+$ is the
phase correlation length for $T > T_{KTB}$ \cite{nelson}. AR
probes the superfluid density at the finite wavenumber $q \sim
1/\xi$ because it occurs within the pairing correlation length, $\xi$,
around the interface between, say, a NS
junction. Hence, AR, being a proximity effect, is not particularly sensitive to
the long wavelength physics, which determines $T_c$, and is expected to be
present above $T_c$ because $\rho(q,T) \sim (\xi_+/\xi )^2 > 0$ for $T^*>T>T_c$.

To provide more quantitative calculations of these qualitative arguments,
we will apply the Bogoliubov-de Gennes
(BdG) equation \cite{degennes} with appropriate boundary
conditions \cite{blonder} to a pseudogap state assumed to be described
in terms of the PIPP \cite{emery}.
As expected, we find that the AR is still present well above
$T_c$ and its effects may be observed in tunneling experiments.
The tunneling conductance, $G=dI/dV$, as a function of the bias
voltage, $V$, measured along the in-plane \{100\} direction of $d$-wave
superconductors for a small tunneling barrier exhibits a conductance enhancement
(CE) inside the superconducting gap \cite{wei,deutscher}. 
It is enhanced by
the factor of 2 in the low $T$ and small tunneling barrier limit
with a edge at the pairing amplitude \cite{tanaka}. We will show that, if the
pseudogap state can be described in terms of the  PIPP, this CE will also be
present above, as well as below, $T_c$.

In a paired state, the quasiparticles (Bogoliubovons)
can be written in terms of a two-component spinor as
$\psi(\bf x) = \left( \begin{array}{c}
f(\bf x) \\ g(\bf x) \end{array}
\right) $, where $f({\bf x})$ and $g({\bf x})$ are, 
respectively, the
electron and hole components and
obey the BdG equation \cite{degennes}. Inspection of the BdG equation
shows that $\psi({\bf x})$ oscillates on a length
scale of $k_F^{-1}$, where $k_F$ is the Fermi wavenumber, so we
introduce the transformation
$\left( \begin{array}{c}
f(\bf x) \\ g(\bf x) \end{array}
\right) = e^{i {\bf k}_F \cdot \bf x}
\left( \begin{array}{c}
u(\bf x) \\ v(\bf x) \end{array}
\right)$, where $u({\bf x})$ and $v({\bf x})$ now vary on a much longer
length scale than $k_F^{-1}$. For $k_F \xi \gg 1$
the BdG equation then becomes \cite{andreev,tanaka,hu}
\begin{eqnarray}
E u({\bf x}) &=& -\frac{i}{m} {\bf k}_F \cdot \nabla u({\bf
x}) + \Delta({\bf k}_F, {\bf x}) v(\bf x) ,
\nonumber \\
E v({\bf x}) &=& \frac{i}{m} {\bf k}_F \cdot \nabla v({\bf x})
+ \Delta^* ({\bf k}_F, {\bf x}) u(\bf x).
\label{bdg}
\end{eqnarray}
We write the pairing order parameter as
$\Delta({\bf k}_F, {\bf x}) = \Delta_k \exp[i \theta({\bf x})]
\Theta(x)$, where $\Delta_k = \Delta_d [\cos(k_x a)
-\cos(k_y a)]$, $a$ the lattice constant of a copper oxide
plane, and $\Theta(x)$ is the step function.
${\bf x} = (x, y, z)$ is a 3D vector, where we
take $x$ to be normal to the interface and $z$ parallel to the
interface and normal to the copper oxide plane. The $x>0$ and $x<0$ sides
correspond, respectively, the paired and normal states. 
In the PIPP scenario, the Cooper pairs are formed above $T_c$, but the
phase fluctuations of the pairing order parameter destroy phase coherence and
hence superconductivity above $T_c$. Such phase fluctuations are taken to be 
thermal and static as in the previous works \cite{emery,franz}.
Other kinds of phase fluctuations such as quantum or temporal ones were argued
to be less important \cite{emery,franz}.
The thermal fluctuations destroying superconductivity above $T_c$ are due to
a plasma of unbound vortices and antivortices, and are described by the 2D $XY$
model \cite{emery}. Each vortex is surrounded by a circulating supercurrent which
is related with the order parameter phase $\theta({\bf x})$ by ${\bf v}_s ({\bf
x}) = \hbar \nabla \theta ({\bf x})/(2m)$, where ${\bf v}_s ({\bf x})$ is the
local superfluid velocity.

Eq.\ (\ref{bdg}) can be solved by taking
\begin{eqnarray}
\left( \begin{array}{c} u(\bf x) \\ v(\bf x) \end{array}
\right) = e^{i {\bf k} \cdot \bf x}
\left( \begin{array}{c} u e^{i\theta/2} \\ v e^{-i\theta/2}
\end{array} \right)
\label{uv}
\end{eqnarray}
to yield a Doppler-shifted local quasiparticle excitation spectrum of
\begin{eqnarray}
E = {\bf k}_F \cdot {\bf v}_s ({\bf x}) +\sqrt{({\bf k}_F \cdot
{\bf k}/m)^2 +\Delta_k^2} .
\label{energy}
\end{eqnarray}
In writing Eq.\ (\ref{energy}), we have assumed that
the phase $\theta ({\bf x})$ varies slowly on the scale of a single pair
size $\xi$ in the sense that it is meaningful to
specify the energy vs. wavevector relation of a quasiparticle at the
position {\bf x}. This condition
can be written as $ {\bf k}_F \cdot {\bf v}_s ({\bf x}) <
\Delta_d$, which is satisfied in the pseudogap state as estimated by Franz
and Millis \cite{franz}, as we will discuss later.          
The change in the local quasiparticle excitation spectrum of Eq.\
(\ref{energy}) will affect the spectral propoerties of the superfluid in that
a physical observable must be averaged over the positions of the
fluctuating vortices. A similar situation arises, as noted by Volovik
\cite{volovik} in the mixed state of a $d$-wave supercondcutor where the
superflow around the field-induced vortices lead to the residual DOS proposional
to $\sqrt{H}$. This contributes $\sim T\sqrt{H}$
to the electronic specific heat, and leads to general scaling relations
\cite{simon}. This suggests that the semiclassical approximation can be
justified in the mixed state. In the present case, we consider a plasma of
thermally induced vortices instead of a regular Abrikosov lattice of
field-induced vortices. The essential physics, however, remains unaltered
because we are interested in the $q \sim 1/\xi$ scale physics.

To determine how the AR will affect physical observables, we
begin with the current, $I$, which
is given by
\begin{eqnarray}
I = Im \left[e \psi^* \nabla \psi \right] = Im \left[e
\left( f^* \nabla f + g^* \nabla g \right) \right].
\label{current1}
\end{eqnarray}
To obtain the quasiparticle wavefunctions $\psi$ of a NS junction,
we follow Blonder $et~al.$
\cite{blonder} and write
\begin{eqnarray}
\psi_N ({\bf x}) = \left( \begin{array}{c} 1 \\ 0 \end{array} \right)
e^{i {\bf k}_e \cdot {\bf x}}
+ A \left( \begin{array}{c} 0 \\ 1 \end{array} \right)
e^{i {\bf k}_h \cdot {\bf x}}
+ B \left( \begin{array}{c} 1 \\ 0 \end{array} \right)
e^{i {\overline{ \bf k}}_e \cdot {\bf x}}
\label{psin}
\end{eqnarray}
for the normal metal, where ${\overline{ \bf k}}_e
= (-k_{ex}, k_{ey}, k_{ez}) $, and
\begin{eqnarray}
\psi_S ({\bf x}) =
C \left( \begin{array}{c} u_+ e^{i\theta/2} \\
v_+ e^{-i\theta/2} \end{array} \right)
e^{i {\bf q}_e \cdot {\bf x}}
+ D \left( \begin{array}{c} v_- e^{i\theta/2} \\
u_- e^{-i\theta/2} \end{array} \right)
e^{i {\overline {\bf q}}_h \cdot {\bf x}}
\label{psis}
\end{eqnarray}
for the paired state.
Here, ${\bf k}_e = (k_F + k ) \widehat{k}_F$, $k=mE/k_F$, ${\bf k}_h =
{\bf k}_F +{\bf k}'$, ${\bf q}_e = {\bf k}_F +{\bf q}$, and
${\bf q}_h = {\bf k}_F +{\bf q}'$.
When the surface is along the \{100\}
direction, one can show from the BdG equation that
$u_+ = u_- = u_0$ and
$v_+ = v_- = v_0$,
where
\begin{eqnarray}
u_0 &=& \left[ \frac{E-\eta +\sqrt{(E-\eta)^2 -\Delta_{\phi}^2}}{2(E-\eta)}
\right]^{1/2}, \nonumber \\
v_0 &=& \left[ \frac{E-\eta -\sqrt{(E-\eta)^2 -\Delta_{\phi}^2}}{2(E-\eta)}
\right]^{1/2}.
\label{uv0}
\end{eqnarray}
Here, $\eta = {\bf k}_F \cdot {\bf v}_s$ and $\Delta_{\phi} = \Delta_d
\cos(2\phi)$ because we consider the \{100\} surface, where $\phi =
\tan^{-1}(k_{Fy}/k_{Fx})$. The energy dispersion relation of Eq.\
(\ref{energy}) determines only the components of the momenta along the
${\bf k}_F$ direction of the incoming electrons such that ${\bf k}'
\cdot {\bf k}_F /m = -E$, $ {\bf q} \cdot {\bf k}_F /m = \sqrt{
(E-\eta)^2 -\Delta_F^2}$, and ${\bf q}' \cdot {\bf k}_F = -{\bf q}
\cdot {\bf k}_F $. The components normal to ${\bf k}_F$ are
determined by the boundary conditions, as can easily be seen.

Applying the boundary conditions at $x=0$,
\begin{eqnarray}
\psi_N (x=0) - \psi_S (x=0) &=& 0,
\nonumber \\
\frac{\partial\psi_S }{\partial x}(x=0) - \frac{\partial\psi_N }
{\partial x}(x=0) &=& 2m V_b \psi_N (x=0),
\end{eqnarray}
where $V_b$ is the barrier
potential energy at the boundary,
we obtain for \{100\} tunneling
\begin{eqnarray}
A &=& \frac{\cos^2 \theta}{D_0} \Gamma, \nonumber \\
B &=& \frac{1}{D_0} \left[ (1-\Gamma^2) \left\{ Z(Z +i \cos \theta) 
+(\wt{\eta}/2)^2 \cos^2 \theta \right\} \right. 
\nonumber \\
&+&  \left. \frac{1}{u_0^2}(\wt{\eta}/2) \cos^2\theta 
\right] , \nonumber \\ 
D_0 &=& \left(1-\Gamma^2 \right) \left[Z^2 +(\wt{\eta}/2)^2 \cos^2 \theta \right]
+\left(1 +\wt{\eta} \right) \cos^2 \theta , \label{coeffi}
\end{eqnarray}
where $\Gamma = v_0/u_0$, with $u_0$ and $v_0$ given by Eq.\
(\ref{uv0}), $\theta$ the tunnelig angle, and $Z$ is the dimensionless barrier
strength given by $Z = m V_b/k_F$. In deriving Eq.\ (\ref{coeffi}), we have made
the approximation $\wt{\eta}_x \equiv mv_{sx}/k_{Fx} \approx \eta/(2\epsilon_F)
\equiv \wt{\eta}$. The average over the tunneling angle was performed
with an equal probability appropriate for small $Z$ \cite{wei}. Inserting Eq.\
(\ref{psin}) into Eq.\ (\ref{current1}), we find that the tunneling current is
given by $I(E) \propto 1 +|A|^2 -|B|^2. $ This current must be averaged over the
phase fluctuations of the order parameter
in the pseudogap state as
$I(V) = \int d \eta P(\eta) I (V, \eta),$
where $P(\eta)$ is the probability distribution of $\eta$ given by
$ P(\eta) = \langle \delta( \eta - {\bf k}_F \cdot {\bf v}_s (\bf
x) \rangle $.
The angular brackets indicate thermodynamic average over the phase
fluctuations governed by the 2D $XY$ model \cite{franz}.
In the cumulant
expansion, $P$ yields a Gaussian distribution of the form
\begin{equation}
P(\eta) = \frac{1}{\sqrt{2 \pi W}} e^{-\eta^2 /(2 W)},
\label{gaussian}
\end{equation}
where $W$ is given by
\begin{equation}
W \approx 3.48 (\alpha_L +\alpha_T) \Delta_d^2 (T/T_c).
\label{fwidth}
\end{equation}
It is crucial to note that $P(\eta)$ was evaluated above $T_{KTB}$ so that we are
in the non-superconducting pseudogap state.
The $\alpha_L$ comes from the longitudinal fluctuations and will
be strongly suppressed in a realistic model
by the Coulomb interaction, while $\alpha_T$ comes from
the transverse fluctuations due to vortices \cite{franz,halperin}.
Franz and Millis found  $\alpha_T \approx 0.1$ \cite{franz}
by fitting Eq.\ (\ref{fwidth}) to the scanning
tunneling spectra \cite{renner}. We use this value below.
%Before we present the detailed calculations of tunneling conductance, a
%qualitative discussion is in order: A PIPP has a
%non-vanishing pairing amplitude of $d$-wave symmetry on the scale of $\xi$,
%and a fluctuating phase $\theta$. 
%If, therefore, $\theta$ varies slowly on the scale
%of $\xi$, or, $\eta < \Delta_d$, the non-vanishing pairing amplitude will be
%felt by incoming quasiparticles, and,
%consequently, CE will show up in tunneling experiments. 
The width of the phase
fluctuation given by Eq.\ (\ref{gaussian}) and (\ref{fwidth}) is $\sqrt{W}
\approx
0.6 \Delta_d$ which means that $\eta < \Delta_d$ is satisfied in the
pseudogap
state of the cuprates. We, therfore, expect that there will exist AR and
residual CE, as shown explicitly below. If, on the other hand, the phase
fluctuations
are too strong in the sense that $\eta \gg \Delta_d$, then the semiclassical
approximation employed to write Eqs.\ (\ref{energy}) and (\ref{coeffi}) is
not valid, and AR will be washed away.

From the current we can calculate the tunneling conductance,
$G=dI/dV$, which after averaging over the phase
fluctuations, is given by
\begin{eqnarray}
G(V) = \frac{1}{\pi} \int_{-\pi/2}^{\pi/2} d\theta \int_{-\infty}^{\infty}
d\eta \frac{1}{\sqrt{2 \pi W}} e^{-\eta^2 /(2 W)} G(V,\eta,\theta),
\nonumber \\
G(V,\eta,\theta) = \frac{e}{\pi} \int_{-\infty}^{\infty} dE
\frac{\partial f(E-V)}{\partial V}[1 +|A|^2 -|B|^2 ],
\label{conduc}
\end{eqnarray}
where $A(E,\eta,\theta)$ and $B(E,\eta,\theta)$ are, respectively, the
coefficients for the Andreev and normal reflections, and are given by Eqs.\
(\ref{uv0}) and (\ref{coeffi}).
In Fig.\ 1, we show the normalized tunneling conductance,
$G(V)/G_n(V)=\pi (1-Z/\sqrt{1+Z^2})^{-1} G(V) $, where $G_n$ is the
tunneling conductance at the normal state for $T>T^*$ averaged over the
angle $\theta$, calculated with
Eqs.\ (\ref{coeffi}) and (\ref{conduc}), as a function of the bias
voltage. The curves are (from above) for $T$ = 1, 1.5, 2, and 3 $T_c$. We take
$W = 0.3~ \Delta_d^2 ( T/T_c )$ and $\Delta_d = 0.04~ eV$ for the
underdoped $T_c = 84$ K BSCCO analyzed by Franz and Millis \cite{franz},
with $\epsilon_F/\Delta_d = 10$.
$Z$ is to be regarded as a fitting parameter to experimental data;
we have taken $Z=0.1$ as a representative value for low barrier tunneling
experiments like the point contact spectroscopies \cite{wei,deutscher}. The
calculated conductance is not very sensitive to the precise parameter values.
Note that, as anticipated, the contribution to the tunneling conductance from
the AR does not vanish upon averaging over the phase
fluctuations because we average $|A|^2$ and $|B|^2$.
In Fig.\ 1, there is a strong CE due to the non-vanishing pairing amplitude,
which is clearly seen up to the highest temperature and arises from the
AR off the PIPP.
At $T=3~T_c$, $G(0)/G_n (0) \approx 1.46$,
for the parameters given
above. Without the normal state pairing, $G(0)/G_n(0)$ would be
equal to 1 above $T_c$.

What is the experimental situation, and what further improvements
in or extensions to the theory should be considered? First, experimentally,
no CE has yet been reported in tunneling experiments above $T_c$.
Our results strongly suggest
a more systematic search for CE for $T>T_c$
in heavily underdoped
and high quality clean samples. In the heavily underdoped
materials, $T_c$ is substantially reduced, which implies that the pseudogap
phenomenon can be probed at the relatively low temperature where the
thermal fluctuations are suppressed. Similarly, high
quality clean samples will also help to observe the CE because the
increased electron mean free path will
enhance the AR \cite{green}.
Second, here as in most previous
calculations, the pairing gap and the
wavefunctions of corresponding quasiparticles were {\it not} determined
self-consistently, because the previous non-self-consistent solutions
of the BdG equation produced satisfactory results for
the tunneling conductances \cite{annett}. 
A main consequence of self-consistent solutions is that the pairing amplitude
varies smoothly from 0 to a bulk value over the
correlation length $\xi$ unlike non-self-consistent case where the pairing
amplitude changes abruptly from 0 to the bulk value at the interface. The phase
of the pairing order parameter, however, remains to be describable in terms of
the 2D $XY$ model, and will be insensitive to a particular way in which the
pairing amplitude is varied, and hence to the self-consistency. It will
neverthless be of interest to study the modifications produced by including
self-consistency. Finally, our result that a CE can be caused by the AR from the
PIPP, which are a particle--particle (pp) condensate, suggests that
one consider possible AR from other forms of condensates. AR is novel in that it
$can$ distinguish pp from particle--hole (ph) condensates, unlike the most
spectroscopic and transport measurements. For a charge density wave (CDW) or
spin density wave (SDW) (both ph condensates), we again anticipate
Andreev-like reflections because an electron is not an elementary excitation
once the condensate is formed. We expect the effects of AR from any ph
condensate to be a {\it dip}, rather than a peak, around zero bias voltage in
the tunneling conductance because the Andreev-reflected particles are
electrons for ph condensate. We are currently exploring this effect and other
potential experimental signatures that may distinguish among models of the
pseudogap state.

To summarize, we have proposed the Andreev reflection as an unambiguous test of the
phase incoherent preformed pair scenario of Emery and Kivelson, which was
substantiated by the quantitative analyses based on the Bogoliubov-de Gennes equation. Experimental
observation of the Andreev reflection above $T_c$, such as the enhanced tunneling conductance
around zero bias voltage along \{100\} direction, could provide a convincing evidence for
the preformed pairs in the pseudogap state. In order to provide more robust argument for our
proposal, we are currently solving, without assuming the semiclassical approximation,
the Bogoliubov-de Gennes equation with the phase
dynamics included fully self-consistently. This will be reported in a separate paper.

We thank Sasha Balatsky, David Pines, Laura Greene, Chandra Varma, Chi Hoon Choi,
and Jaejun Yu
for valuable comments and discussions. We acknowledge support
from KOSEF through grant No.\ 1999-2-114-005-5 (H.Y.C. and Y.K.B.),
the KOSEF 1998
USA Exchange Program (H.Y.C.), and the US NSF under grant No.\ DMR-97-12765
(D.K.C.). H.Y.C. also thanks the Department of Physics, UIUC,
and the CNLS, Los Alamos National Laboratory, for their
hospitality.

\end{multicols}

%\begin{figure}
\vspace{1in}
%\center
%\begin{turn}{-90}
%\epsfig{figure=didv.ps,width=\linewidth}
%\end{turn}

Figure Caption

The normalized tunneling conductance along \{100\} direction as a function of the bias voltage $V$
in units of the pairing amplitude for $Z=0.1$. The curves are, from above on
the $V$ = 0 axis,
for $T$ = 1, 1.5, 2, and 3 $T_c$. The conductance enhancement
due to the Andreev reflections appears
as $G(V=0)/G_n(V=0) > 1$.
%\label{fig1} }
%\end{figure}


\begin{references}

\bibitem{timusk} T. Timusk and B. Statt, Rep. Prog. Phys. {\bf 62}, 61
(1999).
\bibitem{randeria} M. Randeria, in {\it Models and Phenomenology for
Conventional and High-Temperature Superconductivity}, Proceedings of the
International School of Physics ``Enrico Fermi'', Course CXXXVI, edited by
G. Iadonisi, J. R. Schrieffer, and M. L. Chiofalo (IOS Press, Amsterdam, 1998), pp 53-76.
\bibitem{emery} V. J. Emery and S. A. Kivelson, Nature {\bf 374}, 434
(1995).
% E. W. Carlson {\it et al.}, cond-mat/9902077.
\bibitem{levin} Q. Chen, I. Kosztin, B. Janko, and K. Levin, Phys.
Rev. Lett. {\bf 81}, 4708 (1998).
\bibitem{randeria2} J. R. Engelbrecht, A. Nazarenko, M. Randeria, and
E. Dagotto, Phys. Rev. B {\bf 57}, 13406 (1998).
\bibitem{pines} D. Pines, Physica C {\bf 282-287}, 273 (1997).
\bibitem{chubukov} A. V. Chubukov, D. Pines, and B. P. Stojkovic, J.
Phys. Condens. Matter {\bf 8}, 10017 (1996); A. V. Chubukov and J.
Schmalian, Phys. Rev. B {\bf 57}, R11085 (1998).
\bibitem{lee} P. A. Lee and X. G. Wen, Phys. Rev. Lett. {\bf 78},
4111 (1997); P. A. Lee, Physica C {\bf 317-318}, 194 (1999).
\bibitem{anderson} P. W. Anderson, J. Phys.: Condens.
Matter {\bf 8}, 10083 (1996).
\bibitem{orenstein} J. Corson, R. Mallozzi, J. Orenstein, J. N. Eckstein, and
I. Bosovic, Nature {\bf 398}, 221 (1999).
\bibitem{pennington} K. Gorny, O. M. Vyaselev, J. A. Martindale, V. A. Nandor, 
C. H. Pennington, W. L. Hults, J. L. Smith, P. L. Kuhns, A. P. Reyes, and W. G.  Moulton, 
Phys. Rev. Lett. {\bf 82}, 177 (1999).
\bibitem{andreev} A. F. Andreev, Zh. Eksp. Teor. Fiz. {\bf 46},
1823 (1964) [Sov. Phys. JETP {\bf 19}, 1228 (1964)].
\bibitem{bruder} C. Bruder, Phys. Rev. B {\bf 41}, 4017 (1990); C. J.
Lambert and R. Raimondi, J. Phys. Condens. Matter  {\bf 10}, 901, (1998); 
L. Alff, S. Kleefisch, U. School, M. Zittartz, T. Kemen, T. Bauch, A. Marx,
and R. Gross, Eur. Phys. J. B {\bf 5}, 423 (1998).
\bibitem{minnhagen} P. Minnhagen, Rev. Mod. Phys. {\bf 59},
1001 (1987).
\bibitem{nelson} D. R. Nelson, in {\it Phase Transitions and Critical
Phenomena}  edited by C. Domb and J. L. Lebowitz,  (Academic
Press, London, 1983), Vol. 7, p. 1.
\bibitem{degennes} P. G. de Gennes, {\it Superconductivity of Metals
and Alloys}, (Addison-Wesley, New York, 1992).
\bibitem{blonder} G. E. Blonder, M. Tinkham, and T. M. Klapwijk,
Phys. Rev. B {\bf 25}, 4515 (1982).
\bibitem{wei} J. Y. T. Wei, N.-C. Yeh, D. F. Garrigus, and M. Strasik, 
Phys. Rev. Lett. {\bf 81}, 2542 (1998).
%\bibitem{sinha} S. Sinha and K. W. Ng, Phys. Rev. Lett. {\bf 80}, 1296
%(1998);
%M. Covington {\it et al.}, {\it ibid} {\bf 79}, 277 (1997).
\bibitem{deutscher} G. Deutscher, Nature {\bf 397}, 410 (1999).
\bibitem{tanaka} Y. Tanaka and S. Kashiwaya, Phys. Rev. Lett. {\bf 74}, 3451
(1995).
\bibitem{hu} C. R. Hu, Phys. Rev. Lett. {\bf 72}, 1526 (1994); J. H. Xu, J. H.
Miller, and C. S. Ting, Phys. Rev. B {\bf 53}, 3604 (1996).
\bibitem{franz} M. Franz and A. J. Millis, Phys. Rev. B
{\bf 58}, 14572 (1998).
\bibitem{volovik} G. E. Volovik, Pis'ma Zh. Eksp. Teor. Fiz. {\bf 58}, 457 (1993)
[JETP Lett. {\bf 58}, 469 (1993)].
\bibitem{simon} S. H. Simon and P. A. Lee, Phys. Rev. Lett. {\bf 78}, 1548
(1997); N. B. Kopnin and G. E. Volovik,  Pis'ma Zh. Eksp. Teor. Fiz. {\bf 64}, 641 (1996)
[JETP Lett. {\bf 64}, 690 (1996)].
\bibitem{halperin} B. I. Halperin and D. R. Nelson, J. Low Temp. Phys.
{\bf 36}, 599 (1979).
\bibitem{renner} Ch. Renner, B. Revaz, J.-Y. Genoud, K. Kadowaki, and
O. Fischer, Phys. Rev. Lett {\bf 80}, 149 (1998).
\bibitem{green} M. Aprili, M. Covington, E. Paraoanu, B. Niedermeier, and
L. H. Greene, Phys. Rev. B {\bf 57}, R8139 (1998).
\bibitem{annett} A. M. Martin and J. F. Annett, cond-mat/9811389 (unpublished).
% \bibitem{artemenko} S. N. Artemenko and S. V. Remizov, Pis'ma Zh. Eksp.
%Teor. Fiz. {\bf 65}, 50 (1997) [JETP Lett. {\bf 65}, 53 (1997)].
%\bibitem{sinchenko} A. A. Sinchenko {\it et al.}, Pis'ma Zh. Eksp.
%Teor. Fiz. {\bf 64}, 259 (1996) [JETP Lett. {\bf 64}, 285 (1996)].

\end{references}
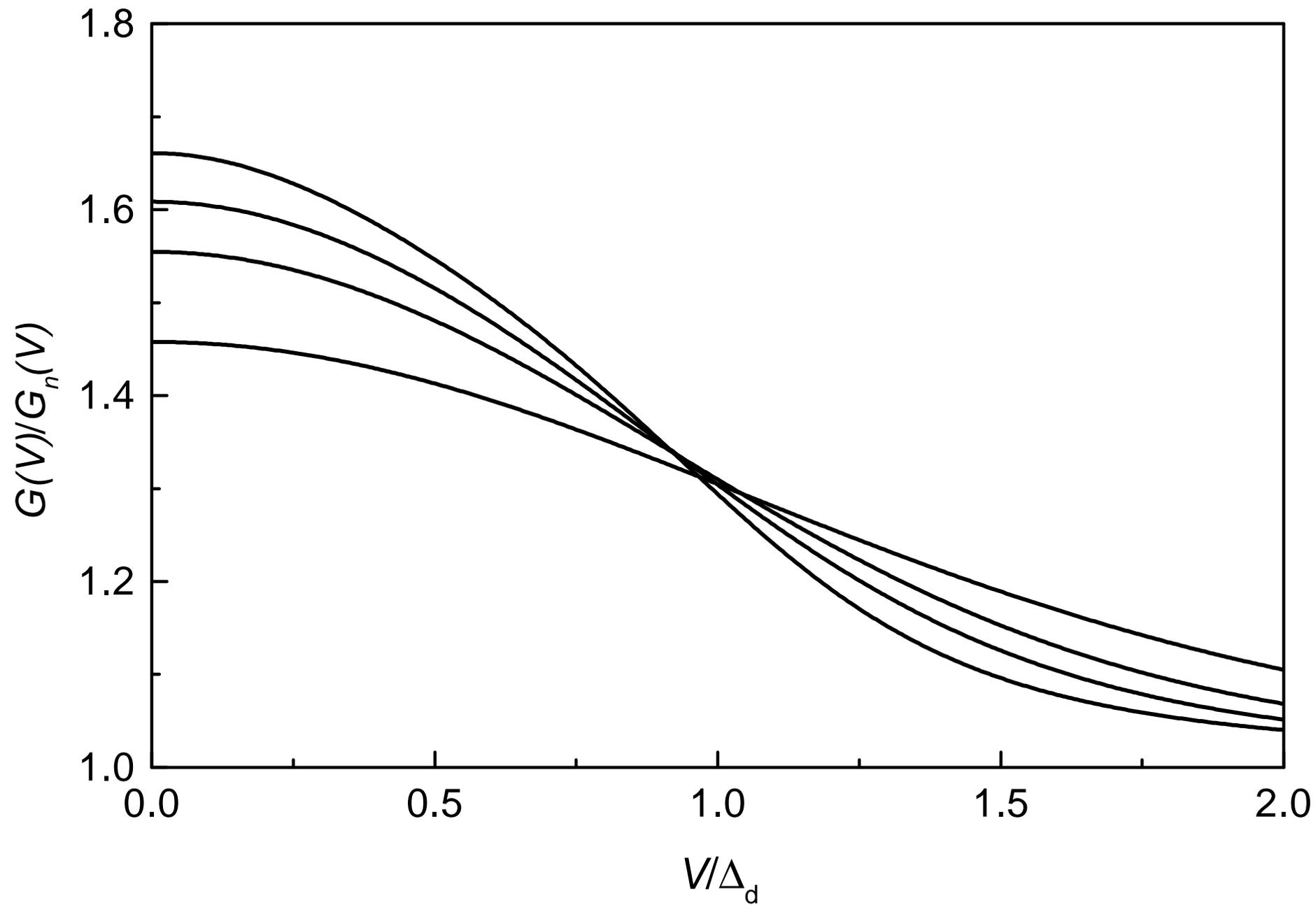
\end{document}